\documentclass[twocolumn,showpacs,amsmath,amssymb,aps,prb,floatfix]{revtex4}

\usepackage{graphicx}% Include figure files
\usepackage{dcolumn}% Align table columns on decimal point
\usepackage{bm}% bold math

\begin{document}

%\preprint{APS/123-QED}

\title{STM/STS Study on $\bm{4a \times 4a}$ Electronic Charge Order and Inhomogeneous Pairing Gap \\in Superconducting $\mathbf{Bi_2Sr_2CaCu_2O_{8+\bm{\delta}}}$}% Force line breaks with \\

\author{A. Hashimoto}
\author{N. Momono}%
\author{M. Oda}
\author{M. Ido}
\affiliation{%
Department of Physics, Hokkaido University, Sapporo 060-0810, Japan
}%

\date{\today}% It is always \today, today,

\begin{abstract}
We performed STM/STS measurements on underdoped Bi2212 crystals with doping levels $p \sim 0.11$, $\sim 0.13$ and $\sim 0.14$ to examine the nature of the nondispersive $4a \times 4a$ charge order in the superconducting state at $T \ll T_c$. The charge order appears conspicuously within the pairing gap, and low doping tends to favor the charge order. We point out the possibility that the $4a \times 4a$ charge order will be dynamical in itself, and pinned down over regions with effective pinning centers. The pinned $4a \times 4a$ charge order is closely related to the spatially inhomogeneous pairing gap structure, which has often been reported in STS measurements on high-$T_c$ cuprates.
\end{abstract}

\pacs{74.25.Jb, 74.50.+r, 74.72.Hs}% PACS, the Physics and Astronomy
%\keywords{STM/STS, pseudogap, electronic charge ordering, 4a$\times$4a superstructure, Bi2212}%Use showkeys class option if keyword
\maketitle

\section{Introduction}

Clarification of the nature of the pseudogap state is expected to provide a clue to understanding the mechanism of high-$T_c$ superconductivity. The pseudogap state appears even in lightly doped regions of $\rm Ca_{2-x}Na_xCuO_2Cl_2$ (Na-CCOC) and $\rm Bi_2Sr_2CaCu_2O_{8+\delta }$ (Bi2212), where the pseudogap is of an asymmetric, V-shaped type at very low temperatures and has been referred to as the zero temperature pseudogap (ZTPG).\cite{Hanaguri,McElroy2005p} It was recently revealed by STM/STS studies that in the ZTPG regime a nondispersive $4a \times 4a$ charge order appears in the two-dimensional (2-d) spatial map of energy-resolved differential tunneling conductance $dI/dV$, which is proportional to the local density of states (LDOS).\cite{Hanaguri,McElroy2005p} A nondispersive $\sim 4a \times 4a$ charge order, electronic in origin, was also reported in the LDOS maps measured in the pseudogap state of Bi2212 at $T > T_c$.\cite{Vershinin} Such a spatial structure in the LDOS maps was first observed around the vortex cores of Bi2212 exhibiting pseudogap-like V-shaped STS spectra with no peaks at the gap edge.\cite{Hoffman2002s,Levy} Such charge orders have attracted much attention because the charge order can be a possible electronic hidden order in the pseudogap state.\cite{Hanaguri,Vershinin}

In measurements of LDOS maps in the superconducting (SC) state of Bi2212, Hoffman {\it et al.} and \mbox{McElroy} {\it et al.} found a strongly dispersive 2-d spatial structure, which has been successfully explained in terms of SC quasiparticle scattering interference.\cite{Hoffman2002s2,McElroy2003n,McElroy2005s} Furthermore, Howald {\it et al.} and Fang {\it et al.} reported a nondispersive $\sim 4a \times 4a$ charge order with anisotropy in the SC state of Bi2212 in addition to weakly dispersive ones, and claimed that the charge order was due to the stripe order and coexisted with the superconductivity.\cite{Howald,Fang} However, the nondispersive $\sim 4a \times 4a$ charge order at $T < T_c$ was not confirmed in other group's LDOS measurements on Bi2212.\cite{Vershinin,Hoffman2002s2} Very recently the nondispersive $4a \times 4a$ charge order at $T < T_c$ was found to appear in heavily underdoped SC Bi2212.\cite{Momono2005j} The charge order is commensurate ($4a \times 4a$) and has an internal structure with a period of $4a/3 \times 4a/3$, which is just like the electronic charge order reported by Hanaguri {\it et al.} for lightly doped Na-CCOC.\cite{Hanaguri} The observation of almost the same charge order for both cuprates Na-CCOC and Bi2212 provides definite evidence that the nondispersive $4a \times 4a$ charge order develops on the Cu-O layer. The $4a \times 4a$ charge order is likely to be dynamical in itself, and pinned down over regions with effective pinning centers.\cite{Momono2005j} To understand the nature of the nondispersive $4a \times 4a$ charge order, it is desirable to investigate the charge order on crystals with different doping levels and/or pinning centers of different natures. 

In the present work, we studied the nondispersive $4a \times 4a$ charge order from STM measurements on Bi2212 crystals with different doping levels and/or pinning centers of different properties. We found that low doping tends to favor the development of the $4a \times 4a$ charge order though it would be dynamical without pinning centers. We also studied the STS spectra over the same region where STM images were taken, and point out that the spatially inhomogeneous gap structure, often reported in STS measurements on high-$T_c$ cuprates,\cite{Pan2001n,Lang,Hoogenboom,Matsuda,Kinoda,MatsudaK,Momono2005} will correlate with the appearance of the pinned $\sim 4a \times 4a$ charge order.

\section{Experimental Procedures}

In the present study, single crystals of Bi2212 were grown using traveling solvent floating zone method. We estimated doping level $p$ of the Cu-O layer from the SC critical temperature $T_c$ determined from the SC diamagnetism and the characteristic temperature $T_{\rm max}$ of the normal-state magnetic susceptibility; both $T_c$ and $T_{\rm max}$ follow empirical functions of $p$.\cite{Oda1990,Nakano1} The doping level was controlled by changing the pressure of oxygen atmosphere in the course of growing the crystal. We performed STM/STS measurements on three different single crystals $\alpha$ ($p \sim 0.11$, $T_c \sim 72$ K), $\beta$ ($p \sim 0.13$, $T_c \sim 78$ K) and $\gamma$ ($p \sim 0.14$, $T_c \sim 81$ K), and report the results on typical sample pairs (A, B), (C, D) and (E, F) cut from $\alpha$, $\beta$ and $\gamma$ single crystals, respectively. In the present STM/STS experiments, Bi2212 crystals were cleaved in an ultrahigh vacuum at $\sim 9$ K just before the approach of the STM tip toward the cleaved surface in situ. Bi2212 crystals are usually cleaved between the upper and lower layers of the Bi-O bilayer. In Bi2212 crystals, excess oxygen atoms contained within Bi-O bilayers provide Cu-O layers nearby with mobile holes. However, excess oxygen atoms will be appreciably lost during the process of cleaving the crystal at high temperatures. So, to suppress the loss of excess oxygen atoms, i.e. mobile holes, to as low a level as possible, we cleaved the crystals at low temperatures ($\sim 9$ K). In the present study, STM images ($512 \times 512$ pixels) were measured over the surface areas of $\sim$ 38 nm $\times$ 38 nm for samples A, C, D, E, $\sim$ 23 nm $\times$ 23 nm for sample B and $\sim$ 19 nm $\times$ 19 nm for sample F in the constant height mode under constant sample-bias voltage $V_{\rm s}$ applied between the tip and the sample. We were able to observe atomically resolved STM images at various bias voltages from a low bias of $V_{\rm s} =$ 10 mV to a high bias of 800 mV. The differential conductance $dI/dV$ was measured by using a standard lock-in technique with an ac bias modulation of 3 mV and a frequency of 4 kHz.

\section{Results and Discussion}

\subsection{STM images of Cu-O layer; the $\bm{4a \times 4a}$ charge order}

\begin{figure}[htbp]
\begin{center}
\includegraphics[width=211pt,clip]{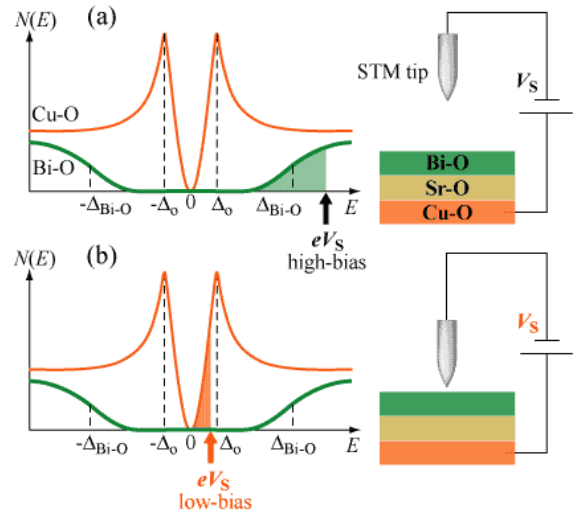}
\caption{(Color online) Schematic energy diagram of Bi2212 and illustration of STM measurements at (a) high ($V_{\rm s} > \Delta _{\rm Bi-O}$) and (b) low biases ($V_{\rm s} < \Delta _{\rm Bi-O}$). The density of states $N(E)$ for Cu-O and Bi-O layers are schematically represented by thin and thick lines, respectively. In the high-bias STM experiment, electron tunneling occurs predominantly between the STM tip and the Bi-O layer when the tip-sample separation is large, whereas in the low-bias STM experiment, it occurs between the STM tip and the Cu-O layer when the tip-sample separation is small.\label{hlbias}}
\end{center}
\end{figure}

\begin{figure}[htbp]
\begin{center}
\includegraphics[width=204pt,clip]{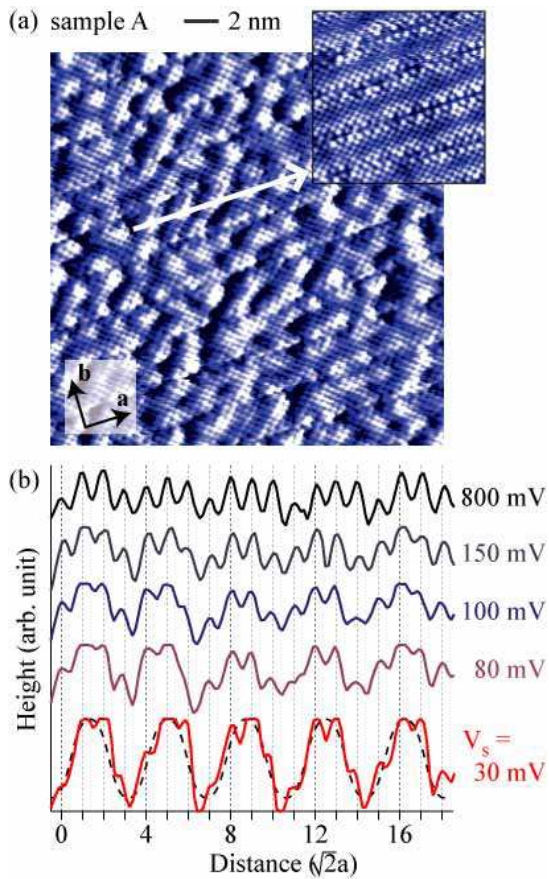}
\caption{(Color online) (a) Part of a low-bias STM image of sample A, measured at a bias voltage of $V_{\rm s} = 30$ mV and an initial tunneling current of $I_{\rm t} = 0.08$ nA at $T \sim 9$ K, showing a $4a \times 4a$ superstructure, together with individual atoms. The inset is part of a high-bias STM image of sample A, measured at $V_{\rm s} = 600$ mV and $I_{\rm t} = 0.3$ nA at $T \sim 9$ K. The image shows a one-dimensional (1-d) superlattice, inherent in the Bi-O layer, with missing atom rows. (b) Line profiles taken along the solid line in the STM image (Fig. 2(a)) for various bias voltages. The solid line is cut perpendicular to the b axis, that is, 45 degrees from the orientation of the $4a \times 4a$ superstructure so that the 1-d superlattice of the Bi-O layer does not obscure the profile of the $4a \times 4a$ superstructure. Note that, in the line profile for the lowest bias, the spatial variation due to the underlying host lattice is partly cut over the intense $4a \times 4a$ superstructure because of saturation of the STM amplifier.\label{stma}}
\end{center}
\end{figure}

The cleaved Bi-O layer of Bi2212 crystals is semiconducting, with a gap of the order of 0.1 eV ($\Delta _{\rm Bi-O}$). Therefore, if we choose a high bias voltage $V_{\rm s}$, which lies outside the semiconducting gap $\Delta _{\rm Bi-O}$ where the electronic states exist in Bi-O layers, in the STM experiment, the STM electron-tunneling occurs predominantly between the STM tip and the cleaved Bi-O layer (Fig.\ \ref{hlbias}(a)). Thus we can observe the Bi-O layer selectively in STM imaging at a high bias ($V_{\rm s}>\Delta _{\rm Bi-O}/$e), when we keep the STM tip at a distance from the sample surface. On the other hand, if we choose a low bias $V_{\rm s}$, which lies within the semiconducting gap $\Delta _{\rm Bi-O}$ where electronic states do not exist in Bi-O layers but in the Cu-O layer , the STM electron-tunneling can occur between the STM tip and the Cu-O layer which is buried just below the cleaved Bi-O layer (Fig.\ \ref{hlbias}(b)). In STM imaging at a low bias ($V_{\rm s}<\Delta _{\rm Bi-O}/$e), we can observe the Cu-O layer selectively when the STM tip approaches the sample surface so that wave functions of carriers between the STM tip and the Cu-O layer can overlap.\cite{Oda1996} In fact, STM images taken on sample A at high and low biases had different features, especially, with respect to the missing atom rows inherent in the Bi-O layer, as shown in Fig. \ref{stma}(a).\cite{Renner1} In the STM image taken at 600 mV (high bias), the missing atom rows appear very clearly (the inset of Fig.\ \ref{stma}(a)). On the other hand, the missing atom rows become very weak in the STM image taken at 30 mV (low bias), as seen in Fig.\ \ref{stma}(a). The latter result confirms that in the low-bias STM imaging the STM tunneling mainly occurs between the STM tip and the Cu-O layer with no missing atom rows.

\begin{figure}[htbp]
\begin{center}
\includegraphics[width=206pt,clip]{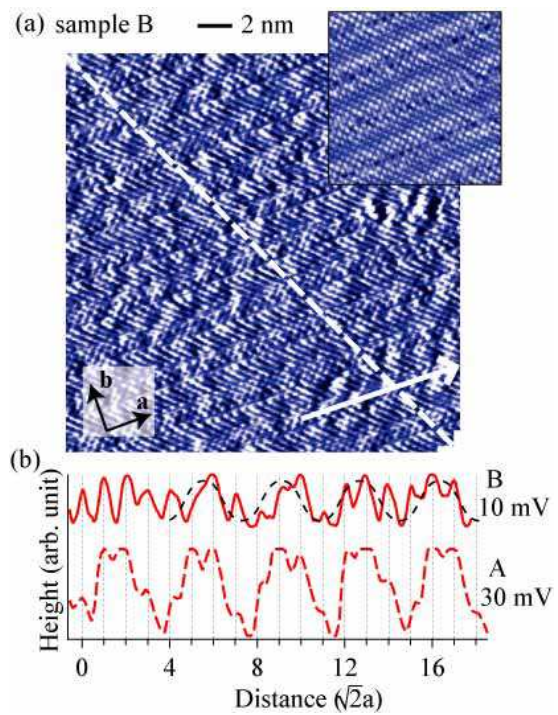}
\caption{(Color online) (a) Part of a low-bias STM image of sample B, measured at $V_{\rm s} = 10$ mV and $I_{\rm t} = 0.08$ nA at $T \sim 9$ K. The inset is part of a high-bias STM image of sample B, measured at $V_{\rm s} = 600$ mV and $I_{\rm t} = 0.3$ nA at $T \sim 9$ K. Note that it shows almost no missing atom rows in contrast with that of sample A (the inset of Fig.\ \ref{stma}(a)). (b) Line profile of STM image taken along the solid line in Fig. 3(a) at $V_{\rm s} =10$ mV (solid line). For comparison, the line profile at $V_{\rm s} = 30$ mV for sample A (Fig.\ \ref{stma}(b)) is also shown (dashed line). The dashed line for sample B is a guide to the eye.\label{stmb}}
\end{center}
\end{figure}

In the low-bias STM image of sample A (Fig.\ \ref{stma}(a)), we can identify a bond-oriented, 2-d superstructure throughout the entire STM image. The 2-d superstructure appeared with the same pattern in both STM measurements at positive and negative biases. Figure \ref{stma}(b) shows the line profiles of STM images, taken along the solid line shown in Fig.\ \ref{stma}(a), for various bias voltages. The 2-d superstructure with a period of $4a$ appears clearly below $V_{\rm s} \sim 100$ mV in addition to the underlying primitive lattice, and the period of $4a$ is almost independent of bias voltage $V_{\rm s}$. The superstructure is more intense at lower biases, while it becomes very weak above $V_{\rm s} = 100$ mV. Part of the low-bias STM image, taken on sample B at $V_{\rm s} = 10$ mV, is shown in Fig.\ \ref{stmb}(a). The 2-d superstructure appears locally on a nanometer scale, not throughout the entire STM image. In Fig.\ \ref{stmb}(b), a line profile of the STM image is shown for the area where the clear 2-d superstructure appears locally and compared with that of sample A. This line profile shows that the period of the superstructure is also $4a$ but its amplitude is much smaller than that observed for sample A.

\begin{figure}[htbp]
\begin{center}
\includegraphics[width=199pt,clip]{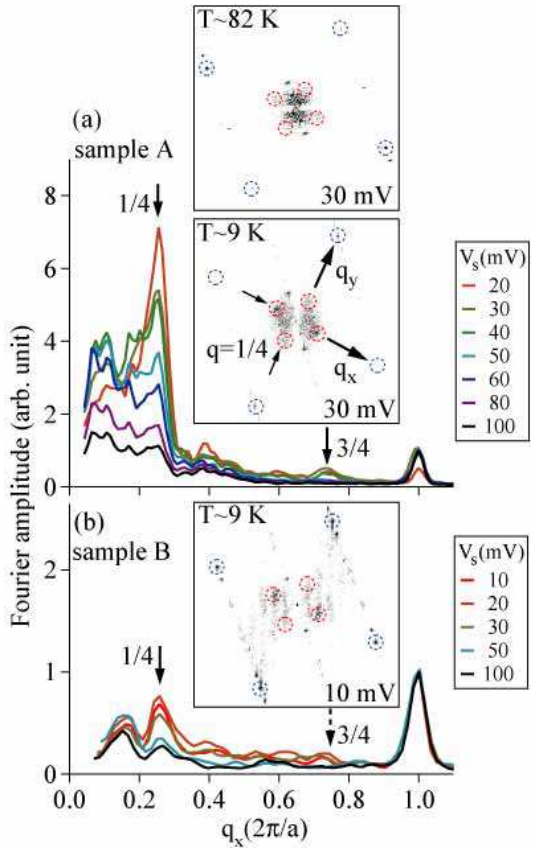}
\caption{(Color) (a) 2-d Fourier maps of the STM images of sample A, taken at $V_{\rm s} = 30$ mV at $T \sim 9$ K (the lower inset) and $\sim 82$ K ($> T_c$) (the upper inset), and cut along the $(0, 0)$-$(\pi, 0)$ line in the Fourier maps at various bias voltages. The Fourier amplitude is normalized by the intensity of the Bragg peak except the amplitude at the lowest bias, which is normalized so that its background level agrees with those for other biases. (b) Fourier map of the STM image of sample B, taken at $V_{\rm s} = 30$ mV at $T \sim 9$ K (the inset), and cut along the $(0, 0)$ - $(\pi, 0)$ line at various bias voltages. The Fourier amplitude is also normalized by the intensity of the Bragg peak.\label{ftab}}
\end{center}
\end{figure}

The 2-d superstructure can also be confirmed in the Fourier map $F(q_x, q_y)$ of the low-bias STM images. The Fourier map $F(q_x, q_y)$ of the STM image, taken on sample A at $V_{\rm s} = 30$ mV, shows that the main Fourier peaks associated with the 2-d superstructure appear at $\bm{q} = (1/4, 0)$ and $(0, 1/4)$, as shown in Fig.\ \ref{ftab}(a). This means that the period of the 2-d superstructure is $4a \times 4a$. The Fourier spot at $\bm{q} = (1/4, 0)$ is stronger than the spot at $\bm{q} = (0, 1/4)$, indicating that the 2-d superstructure is anisotropic. In Fig.\ \ref{ftab}(a), a line cut of the Fourier map along the $(\pi, 0)$ direction is also shown for sample A as a function of the bias voltage $V_{\rm s}$. Each line profile is normalized with the intensity of the Bragg peak at $\bm{q} = (1, 0)$. Weak Fourier peaks also appear at $\bm{q} = (3/4, 0)$ in addition to the strong main peak at $\bm{q} = (1/4, 0)$. Both $\bm{q} = (1/4, 0)$, $(3/4, 0)$ peaks are most intense at the lowest bias (20 mV), but they decrease rapidly with $V_{\rm s}$ and become very weak above $V_{\rm s} \sim 100$ mV. It should be noted here that these Fourier peaks show no change in position and no broadening as $V_{\rm s}$ increases, providing evidence that the present $4a \times 4a$ superstructure is nondispersive. The $4a \times 4a$ superstructure can be observed even above $T_c$ ($= 72$ K) in sample A, as shown in the inset of Fig.\ \ref{ftab}(a).

In Fig.\ \ref{ftab}(b), the Fourier map of the STM image and its line cut along the $(\pi, 0)$ direction are shown for sample B. The $\bm{q} = (1/4, 0)$ Fourier peak appears up to $V_{\rm s} = 50$ mV, with a very weak peak at $\bm{q} = (3/4, 0)$. The intensity of the $\bm{q} = (1/4, 0)$ peak, normalized with the Bragg peak intensity at $\bm{q} = (1, 0)$, is much weaker than that of sample A. The $\bm{q} = (1/4, 0)$ peak decreases with $V_{\rm s}$ and becomes very weak above $V_{\rm s} \sim 50$ mV. In the line cut of the Fourier map for sample B as well as sample A, peak structures are observed at $q < 0.2$. However, the intensity of these peaks is almost independent of $V_{\rm s}$, meaning that these structures are irrelevant to the 2-d superstructure we focused on (Figs. \ref{ftab}(a) and (b)). The present nondispersive $4a \times 4a$ superstructure observed in samples A and B is essentially the same as the nondispersive $4a \times 4a$ charge order with the internal structure of $4a/3 \times 4a/3$ reported by Hanaguri {\it et al.} for lightly doped Na-CCOC.\cite{Hanaguri}

\begin{figure}[htbp]
\begin{center}
\includegraphics[height=417pt,clip]{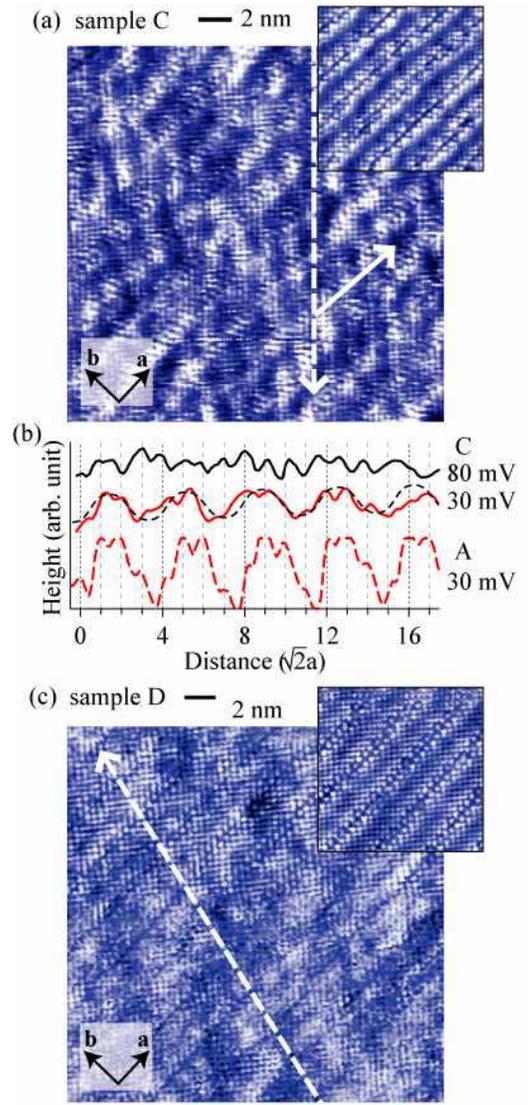}
\caption{(Color online) (a) Part of a low-bias STM image of sample C, measured at $V_{\rm s} = 30$ mV and $I_{\rm t} = 0.09$ nA at $T \sim 9$ K, showing a $4a \times 4a$ superstructure. The inset is part of a high-bias STM image of sample C, measured at $V_{\rm s} = 300$ mV and $I_{\rm t} = 0.3$ nA at $T \sim 9$ K. (b) Line profiles, taken along the solid line in STM image (Fig.\ \ref{stmcd}(a)), at different bias voltages (the solid lines). For comparison, the line profile at $V_{\rm s} = 30$ mV for sample A (Fig.\ \ref{stma}(b)) is also shown (the dashed line). The dashed line for sample C is a guide to the eye. (c) Part of a low-bias STM image of sample D, measured at $V_{\rm s} = 30$ mV and $I_{\rm t} = 0.08$ nA at $T \sim 9$ K, showing a very weak $4a \times 4a$ superstructure. The inset is part of a high-bias STM image of sample D, measured at $V_{\rm s} = 300$ mV and $I_{\rm t} = 0.3$ nA at $T \sim 9$ K.\label{stmcd}}
\end{center}
\end{figure}
\begin{figure}[htbp]
\begin{center}
\includegraphics[width=176pt,clip]{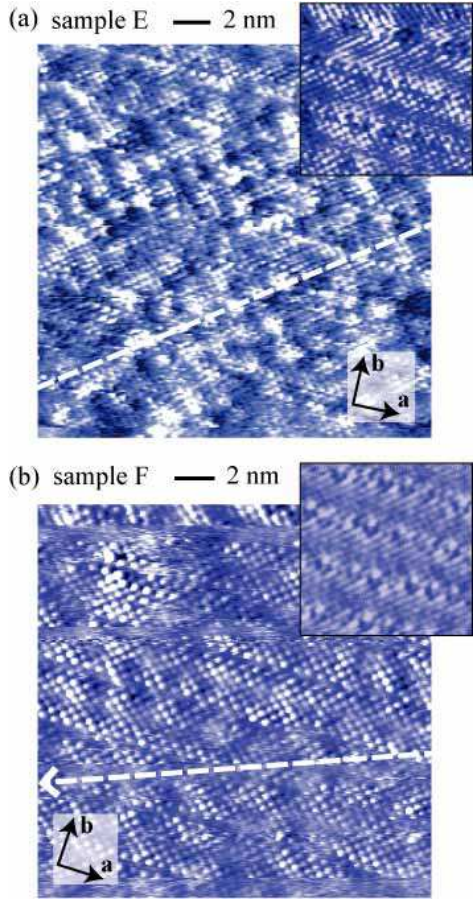}
\caption{(Color online) (a) Part of a low-bias STM image of sample E, measured at $V_{\rm s} = 20$ mV and $I_{\rm t} = 0.08$ nA at $T \sim 9$ K, showing a $4a \times 4a$ superstructure. The inset is part of a high-bias STM image of sample E, measured at $V_{\rm s} = 800$ mV and $I_{\rm t} = 0.3$ nA at $T \sim 9$ K. (b) Part of a low-bias STM image of sample F, measured at $V_{\rm s} = 20$ mV and $I_{\rm t} = 0.08$ nA at $T \sim 9$ K, showing a very weak $4a \times 4a$ superstructure. The inset is part of a high-bias STM image of sample F, measured at $V_{\rm s} = 800$ mV and $I_{\rm t} = 0.3$ nA at $T \sim 9$ K.\label{stmef}}
\end{center}
\end{figure}

Shown in Fig.\ \ref{stmcd}(a) is part of a low-bias STM image taken on sample C at $V_{\rm s} = 30$ mV, which was cut from the single crystal $\beta$ ($p \sim 0.13$, $T_c \sim 78$ K). The 2-d superstructure also appears throughout the entire STM image, as in sample A. Figure \ref{stmcd}(b) shows the line profiles of STM images taken at bias voltages of 30 mV and 80 mV. The superstructure with a period of $\sim 4a$ clearly appears at 30 mV, but its amplitude is much weaker than that observed for sample A. No superstructure appears at a bias of 80 mV. On the other hand, it is difficult to identify 2-d superstructure in the low-bias STM image of sample D, which was cut from the same single crystal ($\beta$) as sample C (Fig.\ \ref{stmcd}(c)). This means that the $4a \times 4a$ superstructure is very weak in sample D. In Fig.\ \ref{stmef}, low-bias STM images are also shown for samples E and F, which were cut from the single crystal $\gamma $ ($p \sim  0.14$, $T_c \sim 81$ K). The 2-d superstructure appears clearly throughout the STM image of sample E, while it is rather weak in the STM image of sample F. 

\begin{figure}[htbp]
\begin{center}
\includegraphics[width=206pt,clip]{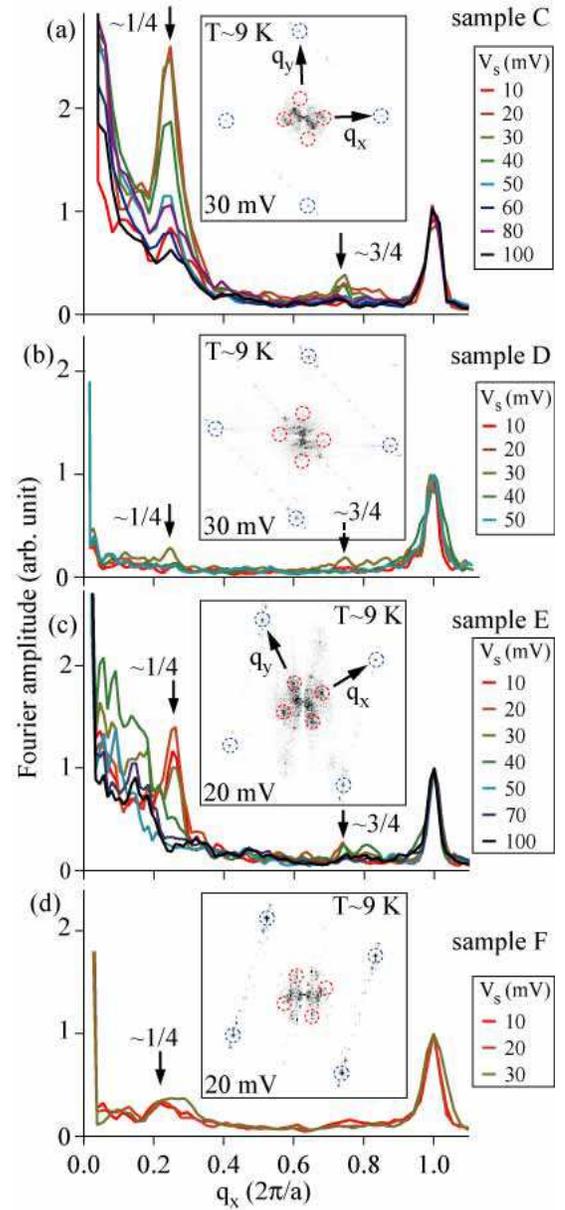}
\caption{(Color) Line cuts of 2-d Fourier maps of STM images along the $(0, 0)$-$(0, \pi)$ lines at various bias voltages for samples C, D, E and F. Here the Fourier amplitude is normalized by the intensity of the Bragg peak. The insets of (a), (b), (c) and (d) are 2-d Fourier maps of STM images shown in Figs.\ \ref{stmcd}(a), \ref{stmcd}(c), \ref{stmef}(a) and \ref{stmef}(b) respectively.\label{ftcde}}
\end{center}
\end{figure}
Shown in the insets of Figs.\ \ref{ftcde}(a) and (c) are the Fourier maps $F(q_x, q_y)$ of STM images taken on samples C and E at $V_{\rm s} = 30$ and $20$ mV respectively. In the Fourier map of sample C, the Fourier transform was carried out except top-right corner of the STM image measured over the area of $38$ nm $\times$ 38 nm, where the main Fourier spot of the 2-d superstructure is split into a few spots because of local distortion of the superstructure. 
The main Fourier spots of the 2-d superstructure appear clearly in samples C and E, as expected from their STM images. The line cuts of the Fourier maps along the $(\pi, 0)$ direction are shown for samples C and E as a function of $V_{\rm s}$ in Figs.\ \ref{ftcde}(a) and (c). The Fourier peaks associated with the 2-d superstructure appear at $\bm{q} = (\sim 0.24\pm 0.01, 0)$, $(\sim 0.74 \pm 0.02, 0)$,
 $(0, \sim 0.24 \pm 0.01)$, $(0, 0.7 \pm 0.02)$ for sample C and $\bm{q} = (\sim 0.26 \pm 0.01, 0)$, $(\sim 0.74 \pm 0.02, 0)$, $(0, \sim 0.24 \pm 0.01)$, $(0, \sim 0.76 \pm 0.02)$ for sample E. The period of the 2-d superstructure of samples C and E is $4a \times 4a$ within experimental error, although we can not rule out the possibility that the superstructure is incommensurate in these samples. 

\begin{figure}[htbp]
\begin{center}
\includegraphics[width=215pt,clip]{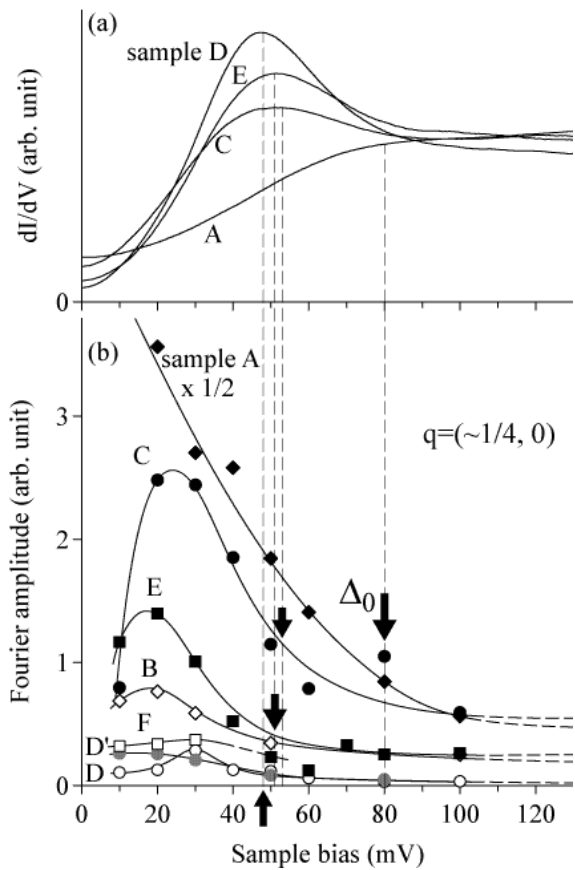}
\caption{(a) Averaged STS spectra of samples A, C, D and E on the positive bias side ($V_{\rm s} \geq 0$). (b) Energy (bias) dependence of the Fourier amplitude at $\bm{q} \sim 1/4$ for sample pairs (A, B), (C, D), (E, F) and sample D$^{\prime}$. The data referred to as samples D and D$^{\prime}$ were measured on different areas of the same cleaved surface. The arrow shows the sample bias $V_{\rm s}$ corresponding to gap size $\Delta _0$ in the averaged STS spectrum shown in Fig.\ \ref{intensity}(a).\label{intensity}}
\end{center}
\end{figure}
The $4a \times 4a$ superstructures of samples C and E are evidently nondispersive, like those of samples A and B, as seen in Figs.\ \ref{ftcde}(a) and (c). It should be noted here that the intensities of the main Fourier peaks associated with the $4a \times 4a$ superstructures of samples C and E show bias voltage dependence different from that of sample A; the peak-intensity is very weak at the lowest bias ($V_{\rm s} = 10$ mV) but rapidly increases with the increase of $V_{\rm s}$ and reaches the maximum at around $V_{\rm s} = 20-30$ mV (Fig.\ \ref{intensity}), where the Fourier peak becomes much stronger than the Bragg peak (Fig.\ \ref{ftcde}(a) and (c)). The intensity of the Fourier peak decreases above $V_{\rm s} \sim 30$ mV and becomes very weak above $V_{\rm s} = 50-60$ mV. 
In Fig.\ \ref{ftcde}(b), the Fourier map of the STM image and its line cuts along the $(\pi, 0)$ direction are shown for sample D. In these line cuts, we can identify the Fourier peak corresponding to the $4a \times 4a$ superstructure only at $V_{\rm s} = 30$ mV, although it is very weak. In Fig.\ \ref{ftcde}(d), the intensity of the main Fourier peak of the $4a \times 4a$ superstructure is also shown for sample F.
 
There exists the possibility that the 2-d superstructures of samples C and E will be incommensurate, as mentioned above. The incommensurate superstructure reminds us of the weakly dispersive $\sim 4a \times 4a$ structure of the LDOS maps which results from the SC quasiparticle scattering interference.\cite{Hoffman2002s2,McElroy2003n}However, the present bias dependences of the wave numbers $\bm{q} = (\sim 0.24, 0)$, $(0, \sim 0.24)$ for sample C and $\bm{q} = (\sim 0.26, 0)$, $(0, \sim 0.24)$ for sample E are too small to be explained in terms of the SC quasiparticle scattering interference (Fig.\ \ref{intensity}(b)).\cite{McElroy2003n} Furthermore, it can hardly be understood in terms of the SC quasiparticle scattering interference that the main Fourier peaks of $4a \times 4a$ superstructure are much stronger than the Bragg peak at $V_{\rm s} = 20-30$ mV.

\subsection{Results of STS; superconducting gap structure and $\bm{4a \times 4a}$ charge order}

\begin{figure}[tbp]
\begin{center}
\includegraphics[width=185pt,clip]{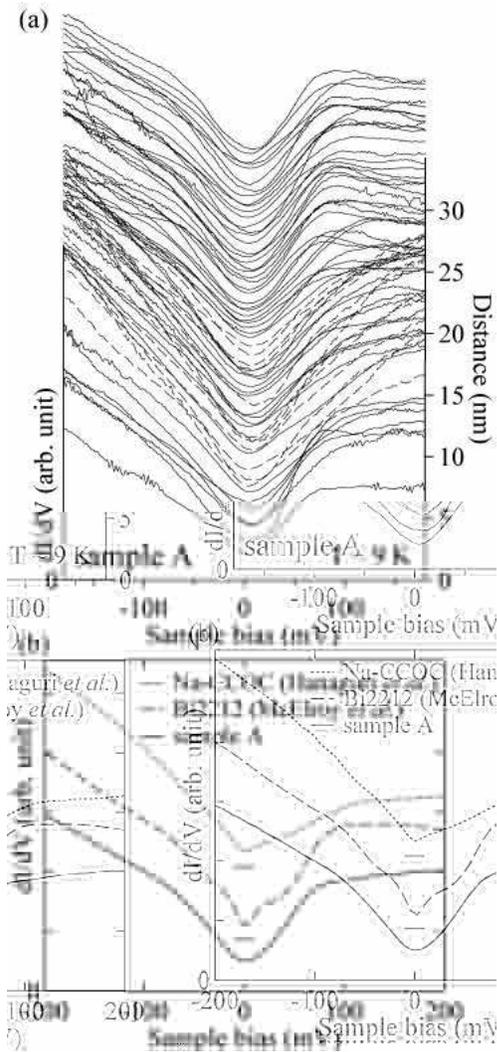}
\caption{(a) Spatial dependence of STS spectra at $T \sim 9$ K for sample A. Solid and dashed lines represent the asymmetric V-shaped ZTPG and the symmetric V-shaped gap with no peaks at the gap edge. (b) STS spectrum averaged over a distance of $\sim 35$ nm on the cleaved surface of sample A at $T \sim 9$ K. Typical ZTPG spectra of lightly doped Na-CCOC (dotted line) and Bi2212 (dashed line) are also shown for comparison.\cite{Hanaguri,McElroy2005p}\label{stsa}}
\end{center}
\end{figure}
\begin{figure}[htbp]
\begin{center}
\includegraphics[width=189pt,clip]{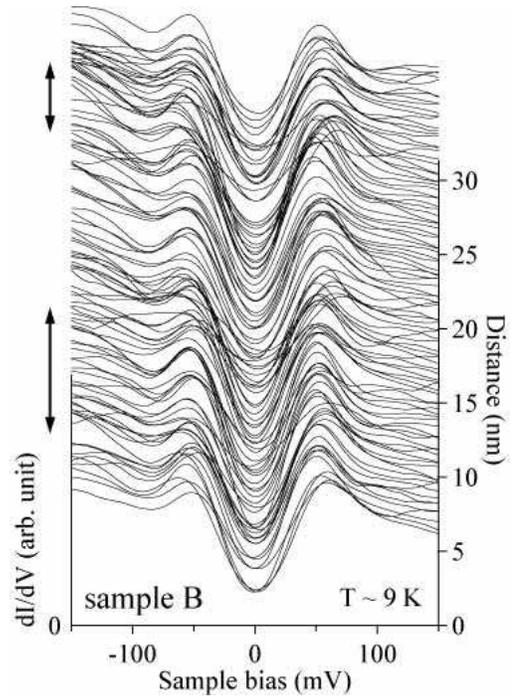}
\caption{Spatial dependence of the STS spectra of sample B, taken at $T \sim 9$ K along the dashed line with a length of $\sim 30$ nm in Fig.\ \ref{stmb}(a). Two-headed arrows beside the spectra indicate the regions where the $4a \times 4a$ superstructure is clearly observed.\label{stsb}}
\end{center}
\end{figure}

Shown in Fig.\ \ref{stsa}(a) is the spatial dependence of STS spectra for sample A, which exhibits an intense $4a \times 4a$ charge order throughout its entire low-bias STM image. Many of the STS spectra show the asymmetric V-shaped ZTPG, but some show a symmetric V-shaped gap with no peaks at the gap edge. Thus, the gap structure of sample A is spatially heterogeneous and inhomogeneous. In Fig.\ \ref{stsa}(b), the spatially averaged spectrum over a distance of $\sim 35$ nm is shown together with the ZTPGs reported for lightly doped Na-CCOC and Bi2212.\cite{Hanaguri,McElroy2005p} The averaged gap structure is very similar to the ZTPGs of Na-CCOC and Bi2212. Width of the averaged gap $\Delta _0$, defined as the width between a shoulder on the positive bias side and zero bias $V_{\rm s} = 0$, is $\sim 80$ meV. In Fig.\ \ref{stsb}, the spatial dependence of STS spectra is shown for sample B, whose low-bias STM image exhibits a weak $4a \times 4a$ superstructure locally. Interestingly the STS spectra of sample B exhibit a homogeneous gap structure of the $d$-wave type. Gap width $\Delta _0$, defined as half of the peak-to-peak width, is $\sim 56$ meV although it tends to be slightly enhanced over the region where the $4a \times 4a$ superstructure appears clearly. 

\begin{figure}[htbp]
\begin{center}
\includegraphics[width=204pt,clip]{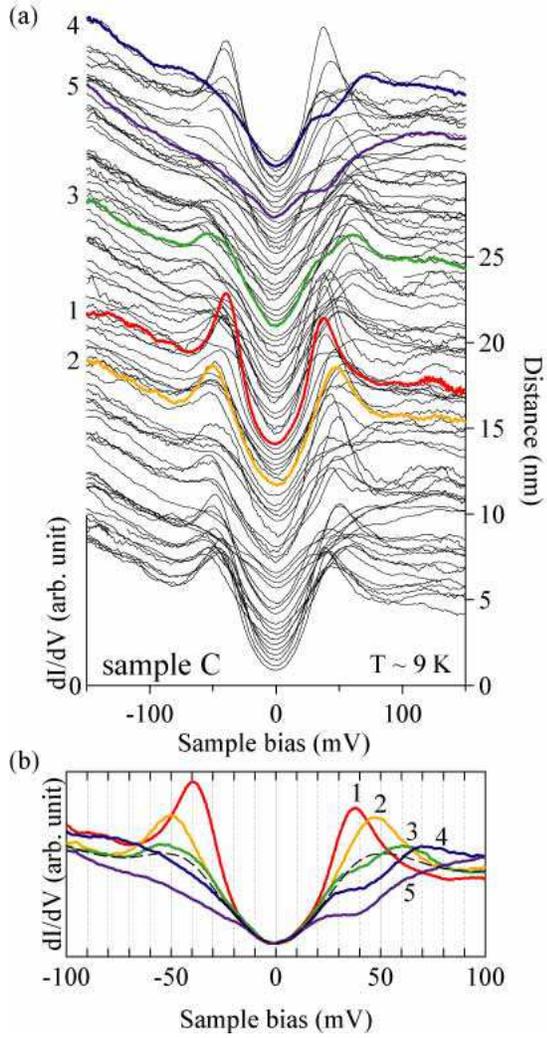}
\caption{(Color) Spatial dependence of STS spectra of sample C, taken along the dashed line with a length of $\sim 30$ nm at $T \sim 9$ K in Fig.\ \ref{stmcd}(a). Representative gap structures are colored. (b) Representative gap structures shown by the colored lines in Fig.\ \ref{stsc}(a). The dashed line shows the STS spectrum averaged over all the STS spectra in Fig.\ \ref{stsc}(a).\label{stsc}}
\end{center}
\end{figure}

\begin{figure}[htbp]
\begin{center}
\includegraphics[width=183pt,clip]{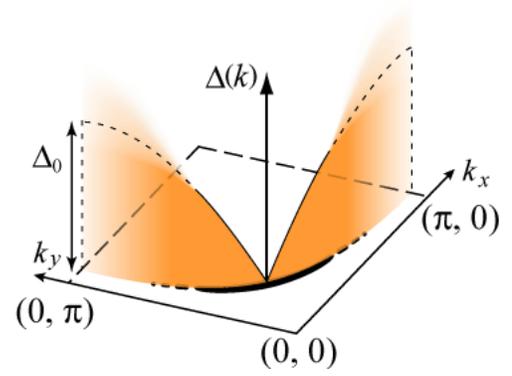}
\caption{(Color online) Illustration for ``the Fermi arc.'' Note that the pairing gap is inhomogeneous around the antinodes near $(\pi, 0)$ and $(0, \pi)$.\label{fermiarc}}
\end{center}
\end{figure}

\begin{figure}[htbp]
\begin{center}
\includegraphics[width=207pt,clip]{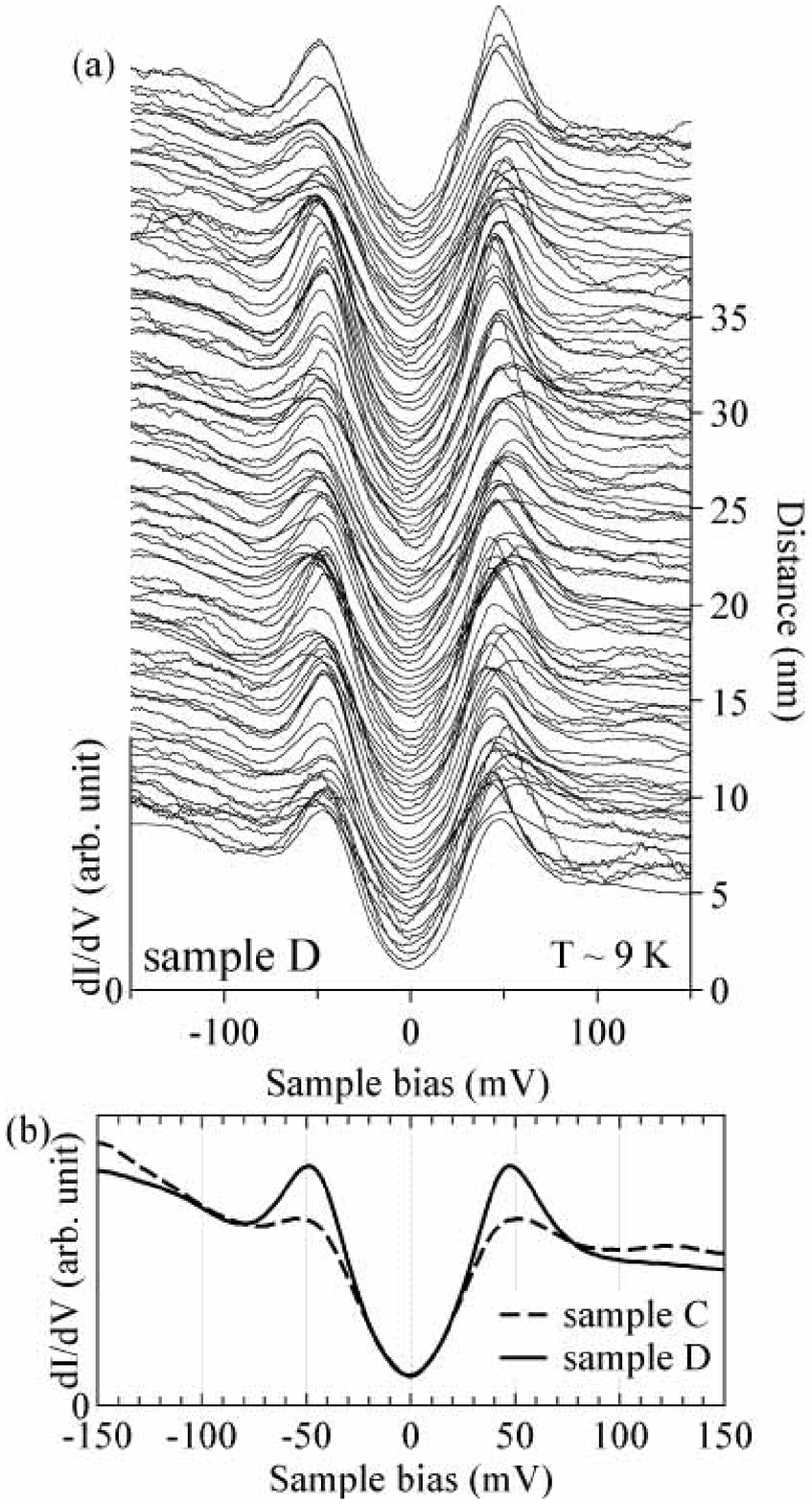}
\caption{(a) Spatial dependence of STS spectra of sample D, taken along the dashed line with a length of $\sim 40$ nm at $T \sim 9$ K in Fig.\ \ref{stmcd}(c). (b) STS spectra averaged over all spectra for samples C and D.\label{stsd}}
\end{center}
\end{figure}

In Fig.\ \ref{stsc}(a), the STS spectra are shown for sample C, which exhibits an intense $4a \times 4a$ superstructure throughout its entire low-bias STM image. It should be stressed that the STS spectra are spatially inhomogeneous. Representative gap structures in STS spectra of Fig.\ \ref{stsc}(a) are shown in Fig.\ \ref{stsc}(b); the gap structure ranges from a typical $d$-wave type to an asymmetric V-shaped type with no peaks at the gap edges, and a gap with larger width tends to be accompanied by a subgap. The variation of gap structure shown in Fig.\ \ref{stsc}(b) is very similar to that reported by McElroy {\it et al.} for underdoped Bi2212 crystals.\cite{McElroy2005p} It should be noted that the gap structure around the bottom is almost the same among all the STS spectra, although the entire gap structure differs among them, as seen in Fig.\ \ref{stsc}(b). This means that the quasiparticle states around the nodes of $d$-wave gap, which dominate the gap structure around the bottom, are homogeneous, and so the inhomogeneity of the gap structure should be attributable to the nature of quasiparticle states away from the nodes, namely, around the antinodes (Fig.\ \ref{fermiarc}). In sharp contrast to sample C, the gap structure of sample D, exhibiting a very weak $4a \times 4a$ superstructure, is spatially homogeneous and of a typical $d$-wave type with $\Delta _0 \sim 48$ meV, as shown in Fig.\ \ref{stsd}(a). The STS data as well as STM data for sample D were obtained over two different areas of the same cleaved surface, but no different features appear in both STS and STM data (Fig.\ \ref{intensity}).
In Fig.\ \ref{stsd}(b), the spatially averaged STS data of samples C and D are shown for comparison. Note that the gap structure around the bottom, $|V_{\rm s}| < 20$ mV, is the same between samples C and D, which were both cut from the same single crystal $\beta$. The agreement means that the pairing gap structure around the node is almost the same in the sample pair C and D, indicating that the doping level is not so different between both samples.

\begin{figure}[htbp]
\begin{center}
\includegraphics[width=200pt,clip]{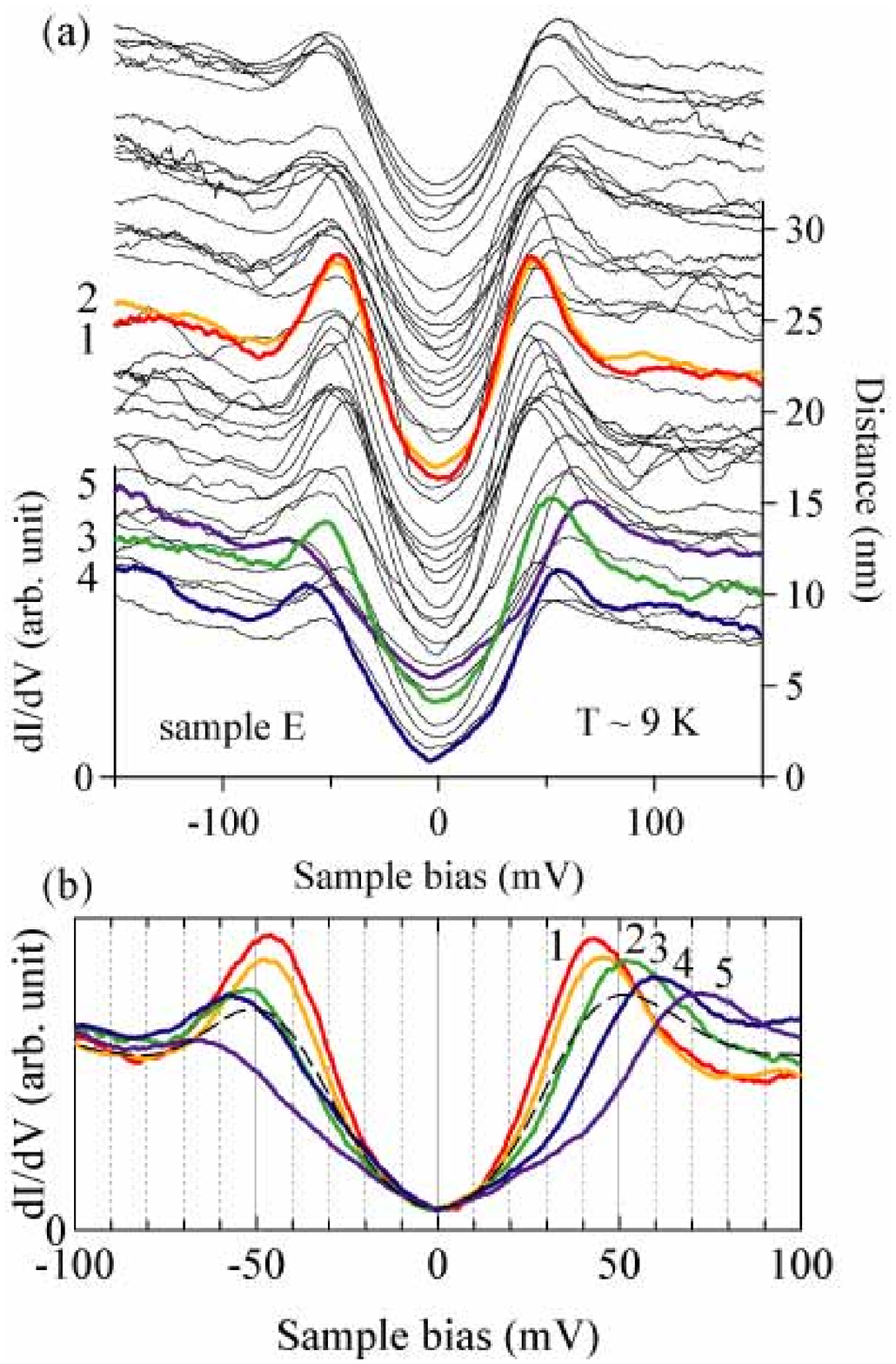}
\caption{(Color) (a) Spatial dependence of STS spectra of sample E, taken along the dashed line with a length of $\sim 30$ nm at $T \sim 9$ K in Fig.\ \ref{stmef}(a). Representative gap structures are colored. (b) Representative gap structures shown by the colored lines in Fig.\ \ref{stse}(a). The dashed line shows the STS spectrum averaged over all the STS spectra in Fig.\ \ref{stse}(a).\label{stse}}
\end{center}
\end{figure}

Figure \ref{stse} shows the STS spectra for sample E, which exhibits a clear $4a \times 4a$ superstructure throughout its low-bias STM images. The STS spectra are also inhomogeneous spatially, as in sample C. 
In Fig.\ \ref{stse}(b) representative STS spectra are shown for sample E. On the other hand, the STS spectra of sample F, exhibiting a very weak $4a \times 4a$ superstructure, is homogeneous as in sample D.

\begin{figure}[htbp]
\begin{center}
\includegraphics[width=196pt,clip]{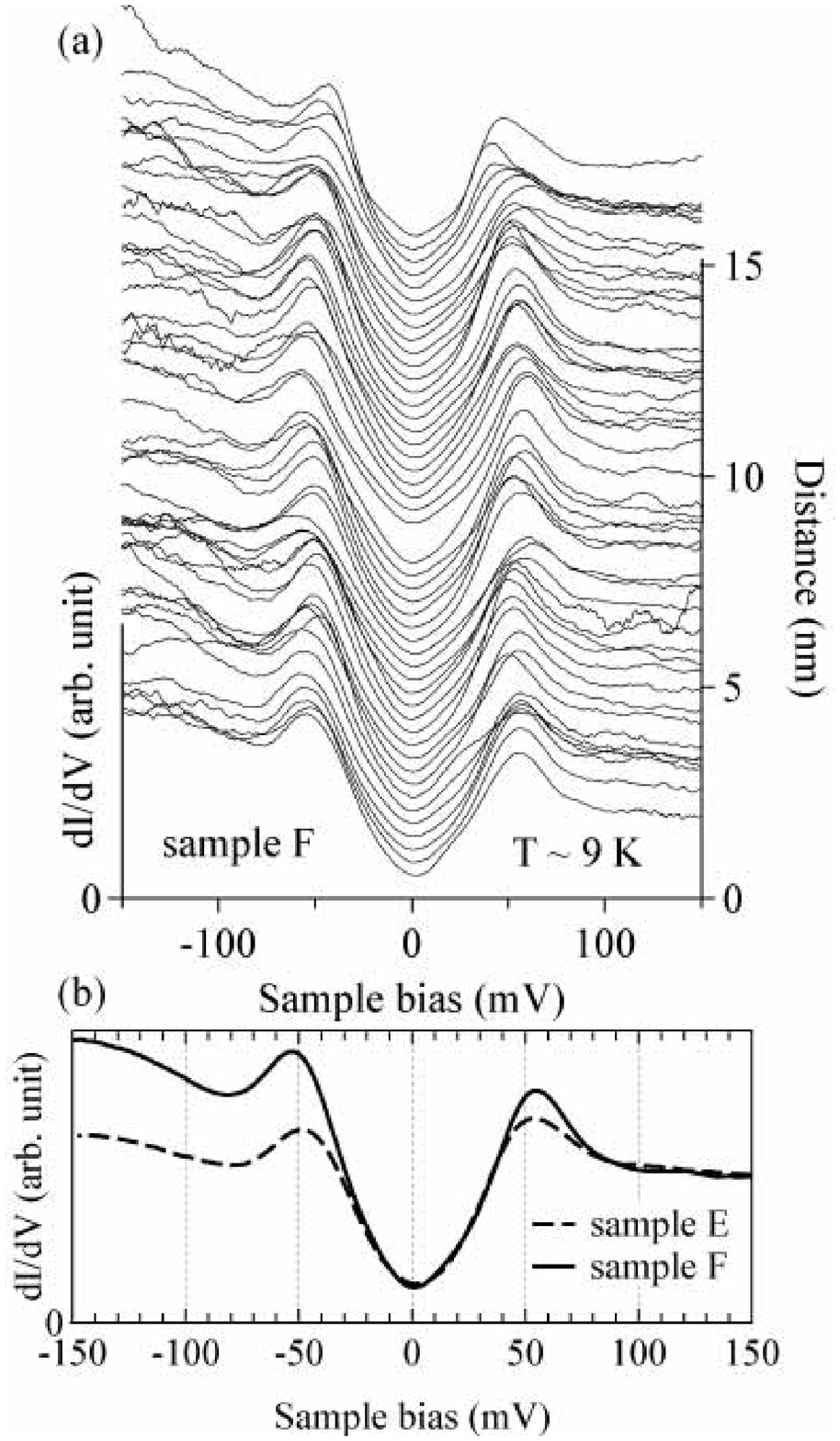}
\caption{(a) Spatial dependence of STS spectra of sample F, taken along the dashed line with a length of $\sim 15$ nm at $T \sim 9$ K in Fig.\ \ref{stmef}(b). (b) STS spectra averaged over all spectra for samples E and F.\label{stsf}}
\end{center}
\end{figure}

It should be emphasized here that STS spectra of samples A, C and E, which exhibit intense $4a \times 4a$ superstructures throughout their low-bias STM images, are spatially inhomogeneous (heterogeneous). 
As seen in Figs.\ \ref{intensity}(b), \ref{stsc}(b) and \ref{stse}(b), sample C exhibits more intense $4a \times 4a$ superstructure than sample E, and the STS spectra of sample C show more various types of the pairing gap than those of sample E, implying that the sample exhibiting more intense $4a \times 4a$ superstructure shows more inhomogeneous gap structure. 
On the other hand, STS spectra of samples B, D and F, exhibiting weak and/or local superstructures, are rather homogeneous. Similar relation between the inhomogeneous gap structure and the development of the $4a \times 4a$ superstructure was confirmed on another sample pair, different from sample pairs (C, D) and (E, F). These facts will lead us to the possibility that inhomogeneous gap structure will be related with the development of the $4a \times 4a$ superstructure. 

In the present study, the STS measurements were carried out on the same area of the cleaved surface where STM images were taken. Thus, the inhomogeneous gap structure, relating with the nature of quasiparticle states around the antinodes, will be intrinsic to the $4a \times 4a$ superstructure state; namely, in the $4a \times 4a$ superstructure state the quasiparticle states around the antinodes will be modified inhomogeneously. McElroy {\it et al.} have reported that the nondispersive $4a \times 4a$ superstructure, caused by a charge order, brings about a severe decoherence effect on quasiparticle states around the antinodes. They claimed that the observation of the charge order would be restricted to outside of the pairing gap with $\Delta _0 \agt 65$ meV.\cite{McElroy2005p} However, this is not the present case. Figure\ \ref{intensity} shows that the $4a \times 4a$ superstructure appears conspicuously within the pairing gap with $\Delta _0 \alt 65$ meV. If the $4a \times 4a$ superstructure is due to some kind of lattice distortion, it is difficult to explain why the superstructure appears within a limited bias (energy) range, especially, associated with the gap size $\Delta _0$; $V_{\rm s} \alt \Delta _0/e$. Such a limitation of $V_{\rm s}$ in observing the $4a \times 4a$ superstructure means that the superstructure is electronic in origin, that is, due to an electronic charge order, as has been claimed in many preceding studies.\cite{Hanaguri,McElroy2005p,Vershinin,Howald,Fang,Momono2005j} The appearance of the $4a \times 4a$ charge order within the pairing gap implies that quasiparticles of the SC state and/or hole pairs will take part in causing the charge order.

\begin{figure}[htbp]
\begin{center}
\includegraphics[width=198pt,clip]{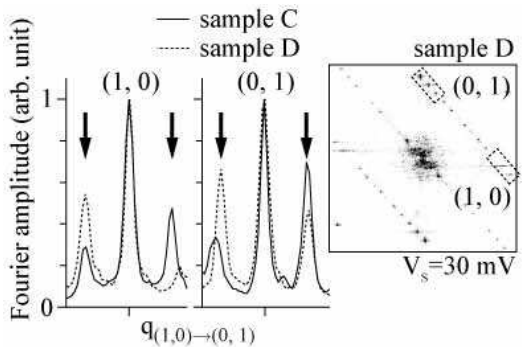}
\caption{Line cuts of the Fourier spots associated with the 1-d superlattice in samples C and D, obtained from the insets of Figs.\ \ref{ftcde}(a) and (b). Here the peak-intensities for samples C and D were normalized with their Bragg-peak intensities. The line cuts were taken along $(1, 0)$-$(0, 1)$ lines around the Bragg spots, as shown by dotted boxes in the inset. The inset is the 2-d Fourier map for sample D shown in Fig.\ \ref{ftcde}(a), although its contrast is enhanced so that the Fourier spots associated with the 1-d superlattice will clearly be visible in the inset\label{superlattice}}
\end{center}
\end{figure}

Here we pay attention to the spatial resolution of STM/STS tip-scanning in order to discuss the interrelation between the development of $4a \times 4a$ charge order and spatial inhomogeneity of the pairing gap. When the resolution of STM tip-scanning depends on wave-vector and happens to be very poor at $q_x$, $q_y \alt 1/4$, the $4a \times 4a$ charge order observed in STM experiment will be smeared and seemingly weakened, although the underlying lattice (each atom) can be observed clearly. In such a case, even if the pairing gap is spatially inhomogeneous in itself, it will appear homogeneous in STS experiments. Because the gap structure will be averaged over a wide area on account of poor spatial resolution of STS tip-scanning. Such a situation might be in samples B, D and F, whose $4a \times 4a$ charge order is weak and gap structure is homogeneous. 
Then, in order to examine whether the spatial resolution of tip-scanning is high enough for observing the $4a \times 4a$ charge order in the present study, we focus on the 1-d superlattice, whose wave vector $\bm{Q}$ points to $(1, 1)$ direction with $|\bm{Q}|\sim 1/(\sqrt{2} \times 5)$. This is because 2-d charge order's wave-vectors $\bm{q} = (\sim 1/4,0)$, $(0, \sim 1/4)$ have a similar component ($|\bm{q}_{\bm{Q}}|\sim 1/(\sqrt{2} \times 4)$) to the 1-d superlattice's wave-vector ($|\bm{Q}| \sim 1/(\sqrt{2} \times 5)$) in the $(1, 1)$-direction. 
As shown in Fig.\ \ref{superlattice}, the FT spots associated with the 1-d superlattice clearly appear not only in sample C exhibiting intense $4a \times 4a$ charge orders but also in sample D exhibiting weak $4a \times 4a$ charge order. Similar results were also obtained for other sample pairs (A, B) and (E, F). These results indicate that there is no essential difference in the spatial resolution of STM/STS tip-scanning ($\bm{Q}$ and $\bm{q}$) among samples A to F. Hence the $4a \times 4a$ superstructures in samples B, D and F will not be weakened by poor spatial resolution of the tip-scanning but intrinsically weak . 

We can also examine the spatial resolution of the tip-scanning from the spatial dependence of the pairing gap. As mentioned above, if spatial resolution of STM tip-scanning is very poor, the spatial dependence of the gap is seemingly homogeneous in STS experiments, even though the gap structure is intrinsically inhomogeneous.
In such a case, the observed homogeneous gap would be essentially the same as the spatially averaged one of inhomogeneous gap data, which could be obtained in STS measurements with good spatial resolution of STS tip-scanning. To check this point, we compare the the spatially averaged gap obtained for sample C with the averaged gap for sample D, because the doping levels of both samples are almost the same, as mentioned above. However, Fig.\ \ref{stsd}(b) shows that the peak structure of the averaged gap is quite different between samples C and D; the peaks of sample D are evidently higher and sharper than those of sample C. The averaged gaps of samples E and F also show a similar tendency (Fig.\ \ref{stsf}(b)). These indicate that the homogeneous gap structures of samples D and F will be intrinsic, not caused by poor spatial resolution of STM/STS tip-scanning.

\subsection{Pinned $\bm{4a \times 4a}$ charge order and inhomogeneous gap structure}

In high-$T_c$ cuprates, since gap width $\Delta _0$ largely depends on doping level $p$, the value of $\Delta _0$ measured by STS provides information about doping level $p$ of Cu-O layers near the cleaved surface.\cite{Oda1997,Miyakawa} Gap width $\Delta _0$ of sample A, exhibiting the intense $4a \times 4a$ charge order, is larger than that of sample B, exhibiting the weak $4a \times 4a$ charge order locally. This means that surface doping level $p$ of sample A is lower than that of sample B, although both samples were cut from the same single crystal $\alpha$. Furthermore, low-bias STM imaging of slightly underdoped Bi2212 with $\Delta _0 \sim 35$ meV exhibits no $4a \times 4a$ charge order, as was previously reported.\cite{Oda1996} These results indicate that low doping tends to favor the development of the $4a \times 4a$ charge order, at least, in the present study.

Doping level $p$ of sample B is lower than that of sample C, because $\Delta _0$ of sample B is larger than that of sample C (Figs.\ \ref{stsb} and \ref{stsc}). Therefore, a more intense $\sim 4a \times 4a$ charge order could be expected to appear in sample B, compared with the charge order of sample C. However, the $4a \times 4a$ charge order of sample B is weak and only appears locally in the low-bias STM image, whereas sample C exhibits an intense charge order throughout its entire low-bias STM image (Figs. \ref{stmb}(a) and \ref{stmcd}(a)). In addition to this result, the $4a \times 4a$ charge order is much weaker in sample D than in sample C, although the doping level is not so different between the samples, as mentioned above. These results indicate the possibility that doping level will not be the only crucial factor necessary for the development of the nondispersive $4a \times 4a$ charge order; there will be some other important factors in addition to the doping level. It should be emphasized here that sample B shows homogeneous STS spectra with a $d$-wave gap, as mentioned above. The specific, homogeneous $d$-wave gap in sample B means that doping level $p$ is rather homogeneous and hole-pairs are uniformly formed throughout Cu-O layers in this sample. These results suggest that the $4a \times 4a$ electronic charge order will develop dynamically throughout Cu-O layers, and it will be pinned down locally over the region with effective pinning centers in sample B. The pinning of the dynamical charge order will enable us to observe it in STM measurements. From the standpoint of this pinning picture, the marked difference of the $4a \times 4a$ charge order between samples C and D, with similar doping levels, can be explained as the difference in the density and/or strength of pinning centers.

The Bi2212 crystals used in the present study belong to the pseudogap regime. In the pseudogap regime, the Fermi surface can be classified into coherent and incoherent parts; the former is centered at the nodal point of the $d$-wave gap and often referred to as ``the Fermi arc'', whereas the latter is around the antinodes, that is, outside the Fermi arcs (Fig.\ \ref{fermiarc}).\cite{Norman1,Yoshida2003,Ronning,Yanase,Pines,Geshkenbein,Furukawa,Wen} This heterogeneous structure of the Fermi surface in the pseudogap regime can provide a possible reason why parts of quasiparticle and/or hole-pair states become inhomogeneous in the intense, pinned $4a \times 4a$ charge order state; the incoherent electronic states around the antinodes, where the pseudogap develops at $T > T_c$, are easily modified by external perturbation caused by the randomness associated with pinning potential of the charge order.

From the standpoint of the present pinning picture, the $4a \times 4a$ charge order can be expected to appear in STM images when the incoherent quasiparticle states around the antinodes (outside the Fermi arcs) contribute to the STM tunneling. In samples with Fermi arcs of a finite size, namely moderately underdoped samples, no $4a \times 4a$ charge order will appear in the STM images at very low biases, where only coherent quasiparticles with very low excitation energy on the Fermi arcs contribute to the STM tunneling (Fig.\ \ref{intensity}). However, the $4a \times 4a$ charge order will appear at higher biases, where incoherent quasiparticles with high excitation energies outside the Fermi arcs contribute to the STM tunneling, as observed experimentally in samples C and E (Fig.\ \ref{intensity}). On the other hand, in heavily underdoped samples A and B with very tiny Fermi arcs, it is plausible that the $4a \times 4a$ charge order appears in STM images even at very low biases. This is because there are incoherent quasiparticle states with very low excitation energies just outside the tiny Fermi arcs, and they contribute to the STM tunneling even at very low biases.

\section{Summary}

We performed low-bias STM imaging on underdoped SC Bi2212 crystals, and confirmed that the nondispersive $4a \times 4a$ electronic charge order appears within the pairing gap at $T < T_c$. The present nondispersive charge order is consistent with the findings in the LDOS maps for the SC state of Bi2212 by Howald {\it et al.} and for the pseudogap state ($T > T_c$) by Vershinin {\it et al.} \cite{Vershinin,Howald,Fang} Howald {\it et al.} have claimed that the nondispersive charge order results from the formation of the stripe order, though this scenario does not so straightforwardly explain why the observation of the $4a \times 4a$ charge order is restricted to within the pairing gap.\cite{Howald,Fang,Kivelson2001,Bosch,Kivelson2003} The appearance of the charge order within the pairing gap is not inconsistent with the models of pair density waves, electronic supersolids, paired-hole Wigner crystallization, or the coexistence of multi-type SC and spin density wave.\cite{Vojta,Ichioka2002,Podolsky2003,Chen,Sachdev,Tesanovic,Franz,Won2005,Ohkawa2005,Fu,Anderson} On the other hand, Vershinin {\it et al.} have claimed that the nondispersive $\sim 4a \times 4a$ charge order at $T > T_c$ is a hidden order of the electron system in the pseudogap state ($T > T_c$).\cite{Vershinin} In that case, the observation that the nondispersive charge order survives even in the SC state means that the hidden order of the pseudogap state will remain essentially unchanged down to below $T_c$.
It is urgently desired to elucidate how the charge order in the pseudogap state ($T > T_c$) evolves into the nondispersive one in the superconducting state.

We pointed out the possibility that the sample dependence of the nondispersive $4a \times 4a$ charge order can be understood qualitatively from the standpoint of the pinning picture, which indicates that the $4a \times 4a$ charge order will be dynamical in itself and pinned down over regions with effective pinning centers. The dynamical $4a \times 4a$ charge order is a possible candidate for the hidden order in the pseudogap regime of pure bulk crystals with no effective pinning centers. We also pointed out that the pairing gap of samples exhibiting more intense, pinned $4a \times 4a$ charge order is spatially more inhomogeneous. The inhomogeneous gap structure can be attributable to incoherent electronic (quasiparticle) states around the antinodes, where the pseudogap develops at $T > T_c$. The electronic (quasiparticle) states will be largely modified there by randomness associated with the pinning potential of the $4a \times 4a$ charge order.

Thanks to useful discussions for Professor F. J. Ohkawa, Professor Z. Te\v{s}anovi\'{c} and Professor K. Maki. This work was supported in part by Grant-in-Aid for Scientific Research and the 21st century COE program ``Topological Science and Technology'' from the Ministry of Education, Culture, Sports, Science and Technology of Japan.

%\bibliography{hppl}

\end{document}